\documentclass[english,pra,preprint]{revtex4}
\usepackage{amsmath,babel}
\usepackage{graphicx}
\newcommand{\dotcup}{\ensuremath{\mathaccent\cdot\cup}}
\usepackage[T1]{fontenc}
\usepackage[latin9]{inputenc}
\DeclareSymbolFont{AMSb}{U}{msb}{m}{n}
\DeclareMathSymbol{\bbH}{\mathbin}{AMSb}{"48}

\begin{document}
  
\title{Entanglement of Periodic States,
the Quantum Fourier Transform and Shor's Factoring Algorithm}
\author{Yonatan Most$^1$, Yishai Shimoni$^{1,2}$ and Ofer Biham$^1$}
\affiliation{$^1$Racah Institute of Physics, 
The Hebrew University, Jerusalem IL-91904, Israel \\
$^2$Department of Neurology, 
Mount Sinai School of Medicine, New York, New York, USA}

\begin{abstract}

The preprocessing stage of Shor's algorithm generates a class
of quantum states referred to as periodic states, on which 
the quantum Fourier transform is applied.
Such states also play an important role in 
other quantum algorithms that rely on the
quantum Fourier transform. 
Since entanglement is believed to be a necessary
resource for quantum computational speedup, 
we analyze the entanglement of periodic states
and the way it is affected by the quantum Fourier transform.
To this end, we  
derive a formula that evaluates the Groverian entanglement 
measure for periodic states. 
Using this formula, we explain the surprising result that the
Groverian entanglement of the periodic states 
built up during the preprocessing stage
is only slightly affected by the
quantum Fourier transform. 

\end{abstract}
  
\pacs{03.67.Lx, 89.70.Cf}

\maketitle
 
\section{Introduction}
  
Quantum algorithms offer a potential speedup over classical 
algorithms in solving a number of problems.
The origin of this speedup is not yet fully understood, 
but quantum entanglement is believed to play a crucial role 
\cite{Ekert1998,Jozsa2003,Vidal2003}. 
Therefore, it is of interest to analyze the entanglement 
of the quantum register during the operation of quantum algorithms
such as Grover's search algorithm and Shor's factoring algorithm 
\cite{Parker2002,Orus2004,Shimoni2004,Shimoni2005,Kendon2006}.
Currently, all
known quantum algorithms presumed to provide an 
exponential speedup over their classical counterparts rely on the
quantum Fourier transform (QFT) 
\cite{Jozsa1998a}. 
The most notable among them is Shor's factoring algorithm 
\cite{Shor1994,Ekert1996a}. 
During the operation of these algorithms, the quantum states of the
register are characterized by multipartite entanglement.
Unlike the case of bipartite entanglement
\cite{Bennett1996a,Wootters1998},
the multipartite entanglement in a register of 
$q>2$ 
qubits,
is not as well understood, partly because no analog of the Schmidt
decomposition was found for multipartite systems. 

In order to evaluate the entanglement of the state of a 
quantum register, an entanglement measure is needed
\cite{Vedral1997,Vedral1998,Vidal2000,Horodecki2000}. 
Axiomatic considerations have provided a set of 
properties that entanglement
measures should satisfy
\cite{Vedral1997,Vedral1998,Vidal2000,Horodecki2000}. 
These properties include the requirement that any entanglement measure
should vanish for product (or separable) states. 
It should be invariant under local unitary operations 
and should not increase as a result of any 
sequence of local operations complemented by
only classical communication between the parties. 
Quantities that satisfy these properties are called 
entanglement monotones. 
These properties provide useful guidelines in the search
for entanglement measures for multipartite quantum states. 
Entanglement measures based on metric properties of the
Hilbert space 
\cite{Vedral1997,Vedral1997a,Vedral1998} 
and on polynomial invariants 
\cite{Barnum2001,Leifer2004} 
were proposed and shown to satisfy these requirements.
Specific measures of multipartite entanglement include the
average bipartite measure
\cite{Emerson2003}
the Groverian measure 
\cite{Biham2002},
and the
geometric measure
\cite{Wei2003}.
A major difficulty in the evaluation of multipartite measures
is that they involve a minimization of a complicated function
in a high-dimensional space.
As a result, there are no general closed-form expressions for these
measures.

The Groverian entanglement generated in Shor's
algorithm was analyzed in  
Ref. \cite{Shimoni2005}. 
It was shown that the entanglement builds up
during the preprocessing stage and that the 
QFT has little effect on the
Groverian measure. 
This is somewhat surprising since, in general,
the QFT operator tends to generate highly entangled
states when it is applied on product states
\cite{Shimoni2005}. 
Furthermore, the superior efficiency of Shor's algorithm is
attributed to the QFT, and since entanglement is considered a
necessary resource for quantum computational speedup, one would
expect that the QFT will induce it. 
It seems as though for the purpose of quantum speedup it suffices 
for the QFT to simply operate on a highly
entangled register rather than generate entanglement by itself.
 
The states generated by the preprocessing stage of
Shor's algorithm are called periodic states.
These states consist of an equal superposition of
basis states whose indices take the form
$i=jr+l$,
where
$j=0,1,2,\dots$,
$r$ is the period and $l$ is referred to as a shift.
It was shown by numerical simulations that
these states have the property of not being further entangled by
the QFT
\cite{Shimoni2005}. 
In this article we explain this surprising property using
an approximated formula for the Groverian entanglement
measure of periodic states. 

The article is organized as follows.
In Sec. II
we present the Groverian measure.
The periodic states generated by the preprocessing stage
of Shor's algorithm are described in Sec. III
and their entanglement is analyzed in Sec. IV.
The effect of the QFT on their entanglement is
considered in Sec. V. The results are discussed
in Sec. VI and summarized in Sec. VII.

\section{The Groverian Entanglement Measure}
\label{sec:groverian}

The Groverian measure of a quantum state 
$|\psi\rangle$ 
of $q$ qubits
is based on the
maximal overlap 
that 
$|\psi\rangle$ 
may have with any product state 
$|\varphi\rangle$,
with the same number of qubits.
The smaller this overlap gets, the more entangled
the quantum state becomes.
We define the square of  
this overlap as

\begin{equation}
P_{\rm max}(\psi) =
\max_{|\varphi\rangle = |\varphi_1\rangle 
\otimes\dots\otimes |\varphi_q\rangle}
\left| \langle\varphi|\psi\rangle \right|^2,
\end{equation}

\noindent
where  
$|\varphi_m\rangle$, 
$m=1,\dots,q$ 
are single qubit states. 
This quantity cannot be decreased by local operations and classical
communication between the parties holding the different qubits. 
Therefore, any nonincreasing function of 
$P_{\rm max}$ 
that vanishes for product states 
(where $P_{\rm max}=1$) 
is a valid entanglement measure. 
Among all these possible measures, 
we have found it useful to use the 
{\it logarithmic Groverian entanglement measure}
\cite{Most2007}

\begin{equation}  
G(\psi) = -\ln \left( P_{\rm max} (\psi) \right),
\end{equation}

\noindent
to which we refer later in this article as the Groverian measure.
This measure has three important advantages over other possible measures:
(i) It is intrinsically a multipartite measure, 
rather than an average
over bipartite measures for different partitions.
(ii) It takes values in the range $[0,\infty)$,
providing a better resolution than measures that are 
limited to the range $[0,1)$.
This is particularly important in the case of highly entangled
states with a large number of qubits.
(iii) This measure is additive in the sense that if the
subsystems $A$ and $B$ are not entangled with each other,
then
$G(\psi_A \otimes \psi_B) = G(\psi_A) + G(\psi_B)$. 
  
The problem with Groverian-type entanglement measures 
(and with many other proposed measures),    
is the difficulty involved in calculating them for general
quantum states. 
The calculation of 
$P_{\rm max}$ 
involves finding the product state
$|\varphi\rangle$ 
for which 
$\left| \langle\varphi|\psi\rangle \right|^2$
is maximal.
The state 
$|\varphi \rangle$
is then referred to as the
nearest product state. 
This is a maximization problem in a high-dimensional
space.
To evaluate the dimensionality of this space, we first consider
the two-dimensional Hilbert space 
$\bbH^2$ 
of a single qubit, which
has four real parameters.
The normalization 
and the insignificance of the global phase
make it possible to express the quantum state of a single
qubit in the form

\begin{equation}
\label{eq:single_qubit}
|\varphi_m\rangle = \sqrt{1-x_m} |0\rangle 
+ \sqrt{x_m} e^{i \theta_m} |1\rangle,
\end{equation}

\noindent
with only two real parameters.
The first parameter, $x_m$, represents the balance between
$|0\rangle$ and $|1\rangle$ in the corresponding qubit,
and takes values in the range $[0,1]$. 
When $x_m=0$ the qubit is in the 
$|0\rangle$ 
state
and when $x_m=1$ it is in the 
$|1\rangle$ 
state.
The second parameter is the relative phase 
$\theta_m$. 
Altogether, finding 
$P_{\rm max}$ 
involves maximizing a suitable function in a  
$2q$-dimensional space.
This function 
typically exhibits a
large number of local maxima.
For a large number of qubits, 
this calculation requires significant computational resources,
except for some special quantum states for which
analytical formulas for $P_{\rm max}$ may be found. 
So far, a general formula is known only for 
two-qubit states (using the Schmidt decomposition) and for a very
restricted set of highly symmetric states that generalize the GHZ
and $W$ states 
\cite{Shimoni2004}. 
Recently, some attempts were 
made to find formulas for 
the entanglement measure of various three-qubit states
\cite{Tamaryan2008,Jung2008}. 
However, so far such formulas were found
only for a restricted set of states. 
For arbitrary states of more than two qubits, 
the Groverian measure can only be calculated numerically.
  
A numerical scheme for calculating the Groverian entanglement is
described in Ref.
\cite{Shimoni2005}. 
In each step of the scheme, one qubit 
$1 \le m_0 \le q$ 
is selected, and the parameter values for all the other qubits in
the product state are fixed. 
The values of 
$x_{m_0}$ 
and
$\theta_{m_0}$ 
for which the overlap with 
$|\psi\rangle$ 
is maximized can then be found analytically. 
Repeating this step for
every qubit several times, a maximum for the overlap over all the
$x_m$'s and $\theta_m$'s is found. 
Using such a formula to
locate the maximum with respect to each qubit 
is much faster than successive
evaluations of the overlap in a steepest descent method.
This is due to the fact that the
evaluation of the overlap requires resources that are exponential in
$q$. 
Such a series of successive jumps in parameter space 
is also less likely to be misled by
local maxima than the steepest descent method.
A slight improvement to
this scheme was used in Ref.
\cite{Most2007}, 
where at each step an
analytical maximization was preformed over two qubits, 
using the Schmidt decomposition. 
This improved scheme is used
in the present work as well. 
Still, the numerical calculation 
of the entanglement is time consuming. 
Furthermore, the lack of an analytical formula makes it
difficult to achieve a better understanding of multipartite
entanglement and its relation to quantum-computational speedup. 
Thus, it is worthwhile to search for analytical formulas 
for the Groverian entanglement of states that are relevant to quantum
algorithms. Such an approximated formula is derived in Sec.
\ref{sec:ent_periodic}.
    
\section{Periodic States in Shor's Algorithm}
\label{sec:periodic}
 
Shor's algorithm aims to find a factor of a given nonprime
integer $N$. This is done by reducing the factorization problem
to the order-finding problem 
\cite{Nielsen2000}. 
In the order-finding problem 
one selects an integer $y$ which is coprime to $N$ and finds 
its order modulo $N$, denoted $r$. 
By recalling that the order of
$y$ modulo $N$ is the smallest integer such that $y^r=1 (\bmod N)$,
one can see that when exponentiating $y^a (\bmod N)$ for
$a=0,1,2,\dots$, the resulting series will be periodic, with a
period $r$. 
This can be done simultaneously for all
values of $a$ by constructing a superposition of the 
$Q=2^q$
computational basis states in a quantum register with $q$ qubits

\begin{equation}
|\psi\rangle = \frac{1}{\sqrt{Q}} \sum_{a=0}^{Q-1} |a\rangle.
\end{equation}  

\noindent
The proper choice of $q$ is described in Ref.
\cite{Shor1994}.
The result of the modular exponentiation can be held in an auxiliary
register:

\begin{equation}
|\psi\rangle = \frac{1}{\sqrt{Q}} \sum_{a=0}^{Q-1} |a\rangle|y^a 
\bmod N\rangle.
\end{equation}

\noindent
Measuring the auxiliary register will randomly select one of its
values, $z=y^l (\bmod N)$ for some $0 \le l < r$, and will also
filter out
from the main register only those values of $a$ for which
$y^a=z$. Since the series $y^a (\bmod N)$ is periodic in $a$ with a
period $r$, the values of $a$ that remain will make out an arithmetic
progression with a common difference $r$, and an initial term $l$. The
main register will then be in a state we refer to as {\it the
periodic state of $q$ qubits, with period $r$ and shift $l$}
(following Ref.~\cite{Jozsa2001}):

\begin{equation}
|\psi^{q}_{r,l}\rangle = \frac{1}{\sqrt{A}} 
\sum_{j=0}^{A-1} |l+jr\rangle ;~~~~~~
A = \left\lceil \frac{Q-l}{r} \right\rceil.
\end{equation}

\noindent  
This ends the preprocessing stage of Shor's algorithm, and here the
QFT is applied.
In analogy to the discrete Fourier transform (DFT), the QFT
is used in order to
reveal periodicities in its input
\cite{Ekert1998}.
In particular, the amplitudes of the
state 
$|\psi^{q}_{r,l}\rangle$ 
make out a periodic series, and when the
DFT is applied to it, the resulting series 
can be approximated 
by a
periodic series of the same sort, that is, one in which the indices of the
nonzero terms make out an arithmetic progression. In the resulting
series, though, the common difference is $Q/r$, the initial term is
zero, and additional phases are added.
This can be seen through the exact
formula for the resulting series $(y_j)_{j=0}^{Q-1}$, 
given by

\begin{equation}
\label{eq:exactqft}
y_j = \frac{1}{\sqrt{QA}}
\frac{\sin\left( \pi jrA/Q \right)}
{\sin\left( \pi jr/Q  \right)}
e^{-\frac{j}{Q} 2 \pi i \left[l+\frac{1}{2} r (A-1)\right]}.
\end{equation}

\begin{figure}
\hspace*{-2.05cm}\includegraphics*[scale=0.7]{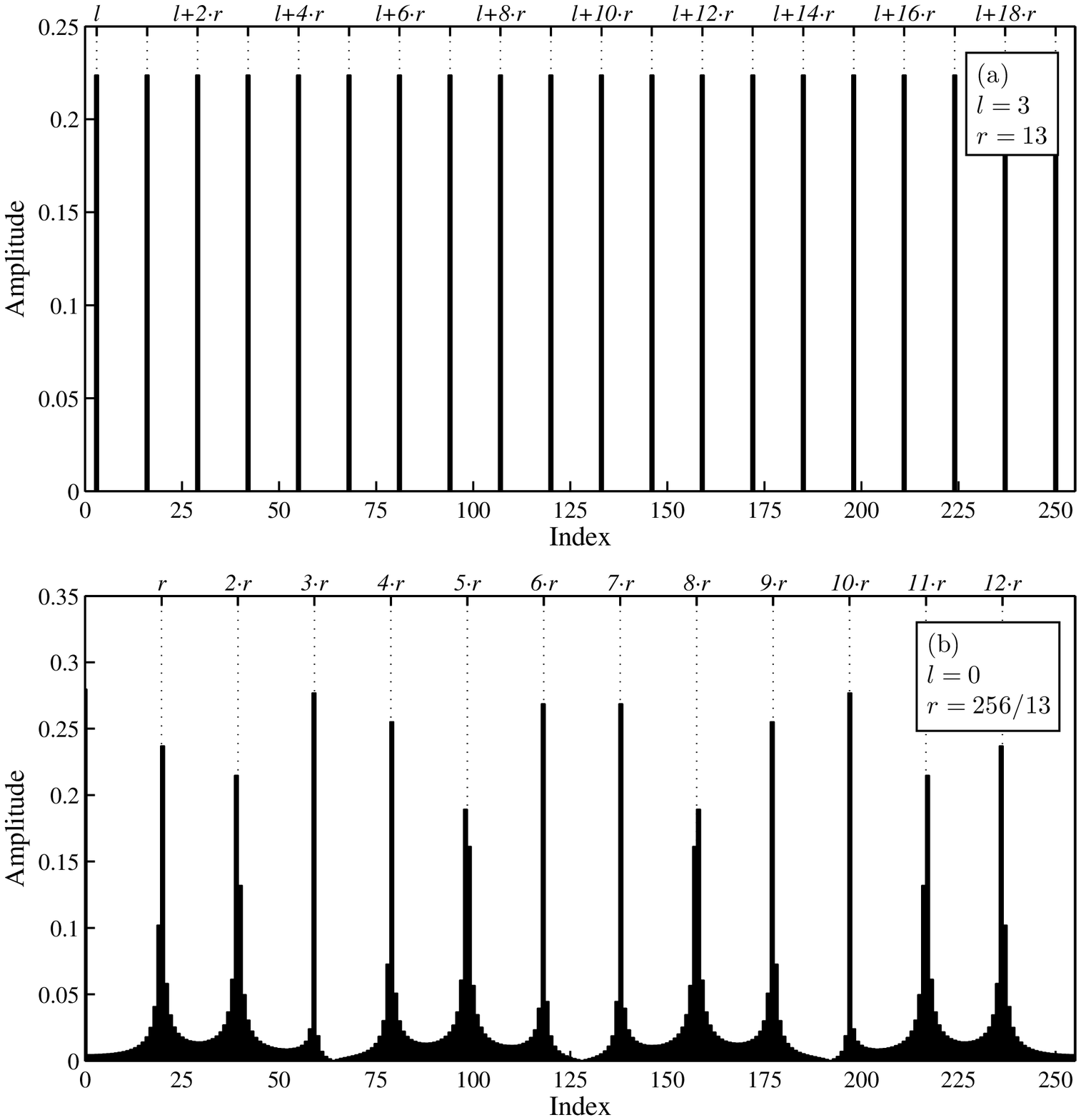}
\caption{
An example of the operation of the QFT on a periodic state. The
periodic state of $q=8$ qubits with period $r=13$ and shift
$l=3$ was transformed by the QFT. The amplitudes of the
original state are given (a), as well as the amplitudes of the
resulting state (b). The dotted vertical lines mark the
multiplicands of the periods of each state 
($Q/r \approx 19.7$).
}
\label{fig1}
\end{figure}

\noindent
Since applying the QFT to a quantum state is equivalent to applying the
DFT to its amplitudes, the action of the QFT on periodic
states can be approximately described as:  

\begin{equation}
\label{eq:qftperiodic}
|\psi^{q}_{r,l}\rangle \xrightarrow{QFT} |\psi^{q}_{Q/r,0}\rangle,
\end{equation}

\noindent
where relative phases are ignored. Within this approximation, the QFT
induces two changes in the periodic state, in analogy with the DFT:
the period is changed from $r$ to $Q/r$, and the shift is changed from
$l$ to $0$ (Fig.~\ref{fig1}).
This removal of the shift is the crucial effect that makes
it possible to extract the period in the next step, 
in which a measurement is performed.
The result of the measurement is not affected by
relative phases; thus, they can be ignored.
The measurement result is an integer close to $jQ/r$
for some $j$, and dividing by $Q$ we are left with a number close to
$j/r$. A continued fraction expansion can then reveal $j$ and $r$. If
we had measured the state before applying the QFT, the result
would have been an integer of the form $l+jr$, which does not allow finding
the period without knowledge of the shift. It is clear that periodic
states are central to the success of Shor's algorithm, and their
importance is further demonstrated by their use in several other
algorithms that make use of the QFT, all of which belong to a class
of problems derived from the hidden subgroup problem
\cite{Jozsa2001}.
  
\section{The Entanglement of Periodic States}
\label{sec:ent_periodic}
\subsection{Equal superposition states}
  
The lack of a general analytical formula for the
Groverian entanglement makes it hard to construct a model that
explains the fact that the QFT does not seem to affect the
entanglement of periodic states. Nevertheless, an approximated formula
may suffice, provided it remains close to the exact value when the number
of qubits increases. Such an approximation may be attainable, since
the set of periodic states is only a restricted set. 

\subsubsection{Definition of the equal superposition states}

Let us first
consider a somewhat less restricted set of states, namely, states that
are superpositions of any number of computational basis states, with
amplitudes that have equal magnitudes and zero phases. Given some
nonempty subset of basis states $S$, we refer to their
superposition as an {\it equal superposition state of $q$ qubits (ES state)},

\begin{equation}
|\psi^q_S\rangle = \frac{1}{\sqrt{ \left| S \right| }} 
\sum_{k \in S}|k\rangle ;~~~~~~
\emptyset\subset S \subseteq \{ 0,1,\dots ,Q-1 \},
\end{equation}

\noindent
where $|S|$ is the size of the set $S$.
Clearly all periodic states are ES states. Some ES states are
nonentangled, like the computational basis states themselves. Another
example is the {\it complete ES state} $|\eta\rangle$, 
which is the superposition of
all basis states and can be written as $|+\rangle^{\otimes q}$,
where
\begin{equation}
|+\rangle =
\frac{1}{\sqrt{2}} \left( |0\rangle+|1\rangle \right).
\end{equation}

\noindent
Other ES states are maximally entangled, such as the GHZ and $W$
states 
\cite{Dur2000}. 
 
Given any ES state, we would like to refer to the binary
representation of each $k \in S$ as $k=j_1 \dots j_q$, where each
$j_m$ is either $0$ or $1$,
and $m=1,\dots,q$ is the index of the qubit. 
This notation in hand, along with the
notation of Eq.~(\ref{eq:single_qubit}), we can write down the
overlap of the ES state with the product state $|\varphi\rangle$ as a
function of the $x_m$'s and $\theta_m$'s. 
It is this {\it overlap function} we
need to maximize in order to find $P_{\rm max}$:

\begin{equation}
\label{eq:function}
f^q_S(x_1,\dots,x_q;\theta_1,\dots\,\theta_q) = 
\langle\varphi|\psi^q_S\rangle = 
\frac{1}{\sqrt{|S|}}
\sum_{j_1 \dots j_q \in S} 
e^{i \sum_{m=1}^{q} j_m\theta_m}
\prod_{m=1}^{q} C^m_{j_m}
\end{equation}

\noindent
where

\begin{equation}
C^m_j= \left\{ \begin{array}{rl} \sqrt{1-x_m} ~~~&~~~ j=0 \\
\sqrt{x_m}   ~~~&~~~ j=1.
\end{array}\right.
\end{equation}

\noindent
To be more precise, we need to maximize $P=|f^q_S|^2$, which is
equivalent to maximizing the magnitude of the complex function
$f^q_S$, ignoring its phase. In fact, we can fix the relative
phases $\theta_m$ to zero, since Eq.~(\ref{eq:function}) then
becomes:

\begin{equation}
\label{eq:fnophase}
f^q_S(x_1,\dots,x_q) = 
\frac{1}{\sqrt{|S|}}
\sum_{j_1 \dots j_q \in S} 
\prod_{m=1}^{q} C^m_{j_m},
\end{equation}

\noindent
which is clearly not smaller in magnitude. This reduces the dimension
of the search space from $2q$ to $q$ (this is actually a special case
of an observation already made in Ref.~\cite{Shimoni2004}). Another
simplification arises in an ES state for which there is one qubit,
$1 \le m \le q$, that all the basis states in the superposition
``agree'' on (that is, $j_m$ is the same for all $k \in S$). This
qubit is not entangled with the rest of the qubits and can be
factored out. We shall call such an ES state {\it reducible}, since
we can fix $x_m=j_m$ and reduce the dimension of the search space by
one.
  
\subsubsection{Examples: Special high symmetry states}

We have so far reduced our problem to finding the values of
$x_1,\dots,x_q$ for the nearest product state $|\varphi\rangle$, to a
$q$-qubit nonreducible ES state $|\psi^q_S\rangle$, by maximizing
the overlap function $f^q_S$ in Eq.~(\ref{eq:fnophase}). Let us
consider two special cases of this problem, which were already analyzed
in Ref.~\cite{Shimoni2004}. The first one is the $q$-qubit GHZ state:

\begin{equation}
|{\rm GHZ}\rangle = \frac{ |0\dots0\rangle + |1\dots1\rangle }{\sqrt{2}}.
\end{equation}

\noindent
For the GHZ state, the nearest product state is any one of the two
basis states that comprise it:

\begin{equation}
|\varphi\rangle = |0\dots0\rangle  ~~~~~{\rm or}~~~~~
|\varphi\rangle = |1\dots1\rangle,
\end{equation}

\noindent	
and for both states: 
$ P_{\rm max}({\rm GHZ}) = 1/2 $.
The second special case is the $2n$-qubit balanced generalized $W$
state, denoted by $|\phi(n,2n)\rangle$. 
It consists of all the
basis states of $2n$ qubits that have $n$ zeros and $n$ ones:

\begin{equation}
|\phi(n,2n)\rangle = 
{\binom{2n}{n}}^{-\frac{1}{2}} 
\sum_{\sum_{m=1}^{2n} j_m = n} |j_1 \dots j_{2n} \rangle.
\end{equation}

\noindent
In this case the nearest product state is the complete ES state

\begin{equation}
|\varphi\rangle = |\eta\rangle, 
\end{equation}

\noindent
and $P_{\rm max}$ is given by

\begin{equation}
P_{\rm max}[\phi(n,2n)]
= 2^{-2n} {\binom{2n}{n}} \approx \frac{1}{\sqrt{\pi n}}.
\end{equation}

\subsubsection{The approximated formula}

Given a general ES state $|\psi^q_S\rangle$, our aim is to find
the nearest product states $|\varphi\rangle$.
The results presented previously 
motivate us to examine two types
of product states as candidate states:

(i) $|\varphi\rangle = |k\rangle$ for some $k \in S$, which means that each $x_m$
is either $0$ or $1$. This yields $ P^q_S = 1 / | S | $,
where $P^q_S$ is the estimated value of $P$.

(ii) $|\varphi\rangle = |\eta\rangle$ (the complete ES state),
which means that $x_m=1/2$ for all $m$'s. This yields
$ P^q_S = | S | / Q $.

Clearly, the first guess is better for a small $S$, and the second
guess is better for a large $S$. They become equally good for
$|S|=\sqrt{Q}$. Thus, we can combine them into one improved guess:

\begin{equation}
P^q_S=\left\{ \begin{array}{rl}
\frac{1}{\left| S \right|} ~~~&~~~ |S| \le \sqrt{Q} \\
\frac{\left| S \right|}{Q} ~~~&~~~ |S| >   \sqrt{Q}
\end{array}\right.
\end{equation}

\noindent
We note that $P^q_S$ is a lower bound on 
$P_{\rm max}(\psi^q_S)$. 
We can also present a crude argument to support the
claim that $P^q_S$ is a good approximation for 
$P_{\rm max}(\psi^q_S)$. 
For a general product state $|\varphi\rangle$,
consider expanding it to a superposition of basis states. 
In this expansion, we would like to maximize the number of basis states
$|k\rangle$ that have corresponding basis 
states in $|\psi^q_S\rangle$ 
(namely, for which $k \in S$). 
The more such basis states,
the larger the overlap will be (ignoring, for now, the amplitudes of
the states).  In the case of a small $S$, a single basis state is a
good guess for $|\varphi\rangle$, since trying to vary any of the
$x_m$'s away from the edges of their range will add a lot of basis
states to the expansion, and most of them will not be members of
$S$, consequently decreasing $P$. In the case of a large $S$, a
product state is desired with a lot of basis states in its
expansion, since there are a lot of members in $S$, and in that case the
complete ES state is hard to beat.
  
\subsubsection{A counterexample}

It turns out that the approximated formula presented above is not valid
for all the ES states.
A state that violates this formula was analyzed in
Ref.~\cite{Shimoni2004}. 
This is the $q$-qubit $W$
state, which consists of all the basis states that have $q-1$
zeros and a single 1:

\begin{equation}
|W\rangle = \frac{1}{\sqrt{q}} \sum_{m=0}^{q-1} |2^{m}\rangle.
\end{equation}

\noindent
Since for the simple $W$ state $|S|=q$, and $q \le \sqrt{Q}$ for a large
enough $q$ (practically $q \ge 4$ is sufficient), our guess yields one of
the basis states as the nearest product state and
$P^q_S=1/q$. 
However, it turns out that the real nearest product state is

\begin{equation}
|\varphi\rangle = \left( \sqrt{\frac{q-1}{q}} |0\rangle +
\sqrt{\frac{1}{q}}   |1\rangle \right)^{\otimes q},
\end{equation}

\noindent
which yields $P_{\rm max}(W)=\left[ (q-1)/q \right] ^{q-1}$. This
is not only different from $P^q_S$; it is also asymptotically
different, as

\begin{equation}
\begin{array}{l}
\displaystyle  P^q_S \xrightarrow[q \rightarrow \infty]{} 0 \\
\displaystyle  P_{\rm max}(W) \xrightarrow[q \rightarrow \infty]{} \frac{1}{e}.
\end{array}
\end{equation}

\noindent
What property of the simple $W$ state makes it violate the validity of the
approximation? Note that for each qubit, the GHZ and balanced
generalized $W$ states have an equal number of zeros and ones across
all the basis states. The simple $W$ state obviously does not have
this property. To understand why this is important, let us
reconsider the maximization problem of finding the nearest product
state $|\varphi\rangle$.
  
We can explore the possible product states by taking a small variation
around some basis state $|k_0\rangle$. 
For two binary numbers
$k_1$ and $k_2$, let us denote the 
{\it Hamming distance} 
between them, which is the number of bits they
differ on, as $d(k_1,k_2)$. 
We can then divide the set $S$ to disjoint
subsets according to the Hamming distance from $k_0$:

\begin{equation}
\begin{array}{l}
\displaystyle S = S_0 \dotcup \dots \dotcup S_q \\
\displaystyle S_m = \left\{ k \in S ~:~ d(k,k_0)=m \right\}.
\end{array}
\end{equation}

\noindent
Without loss of generality we can fix $k_0=0$ 
(this can be arranged
by applying local {\it NOT} gates which do not affect the entanglement
nor the Hamming distances). 
The Hamming distance $d(k,k_0)$ is
then equal to the number of ones
in $k$, and a small variation around $k_0$ means that the $x_m$'s are
small. The terms in Eq.~(\ref{eq:fnophase}) can then be grouped
according to the subsets of $S$:

\begin{equation}
f^q_S(x_1,\dots,x_q) = \frac{1}{\sqrt{|S|}}
\sum_{n=0}^q \sum_{ \stackrel{j_1 \dots j_q \in S_n}
{j_{m_1},\dots,j_{m_n}=1} } 
\sqrt{x_{m_1}}\cdot\dots\cdot\sqrt{x_{m_n}} 
\prod_{m \neq m_i } \sqrt{1-x_m}.
\end{equation}

\noindent
The $n$th term in this expansion has $n$ multiplicands of the form
$\sqrt{x_m}$, so it is dominant in respect to the $(n+1)$th term. 
Through this expansion we see that the nearest product state is in the close
surrounding of $|k_0\rangle$ only if $S$ contains a lot of terms within a
small Hamming distance from $k_0$. In the case of the simple $W$
state, all the terms in $S$ have a Hamming distance of 1 from
$k_0=0$, thus maximizing the term $n=1$ in the expansion. This shows
that a large value of the function $f^q_S$ can be obtained in the
proximity of the state $|0\rangle$, and indeed this is the case. 

The case of the balanced generalized $W$
state $|\phi(n,2n)\rangle$ is different.
In this case, for each basis state, there are $n^2$
basis states at a Hamming distance of 2, and one basis 
state at the maximal Hamming distance of $2n$. 
In analogy, the GHZ state has a maximal Hamming distance between
its two basis states. 
In both states there is no specific choice of $k_0$ for which all the
basis states are within a small Hamming distance from it. This
observation suggests that in general, for ES states that include
basis states with large Hamming distances from each other, taking a
small variation around some basis state does not aid in finding
the nearest product state.
  
\subsection{Application to periodic states}

Returning to periodic states, the periodic state
$|\psi^{q}_{r,l}\rangle$ 
is an ES state with:

\begin{equation}
S = \{l,l+r,\dots,l+(A-1)r\} ;~~~~~~ A = |S| 
= \left\lceil \frac{Q-l}{r} \right\rceil .
\end{equation}

\noindent
The overlap function in Eq.~(\ref{eq:fnophase}) for this special
case will be denoted by

\begin{equation}
\label{eq:periodic_overlap}
f^q_{r,l}(x_1,\dots,x_q) = 
\frac{1}{\sqrt{A}}
\sum_{j_1 \dots j_q \in S} 
\prod_{m=1}^{q} C^m_{j_m}.
\end{equation}

\noindent
The approximated formula for $P_{\rm max}$, 
which is the square of
the maximum of the preceding function, becomes

\begin{equation}
\label{eq:pmaxperiodic}
P^q_{r,l}=\left\{ \begin{array}{rl}
\frac{1}{A} ~~~&~~~ A \le \sqrt{Q} \\
\frac{A}{Q} ~~~&~~~ A >   \sqrt{Q}
\end{array}\right.
\end{equation}

\noindent
Approximating $A \approx Q/r$, we reach a simpler formula
for $G^q_r$, which is an approximation of the logarithmic Groverian
entanglement of periodic states 
$G(\psi^q_{r,l})$:

\begin{equation}
\label{eq:Gperiodic}
G^q_r=\left\{ \begin{array}{rl}
-\ln\frac{1}{r} ~~~&~~~ r <   \sqrt{Q} \\
-\ln\frac{r}{Q} ~~~&~~~ r \ge \sqrt{Q}      
\end{array}\right.
\end{equation}

\begin{figure}
\setlength{\abovecaptionskip}{3pt}
\hspace*{-1cm}\includegraphics*[scale=0.9,trim=0 13 0 17]{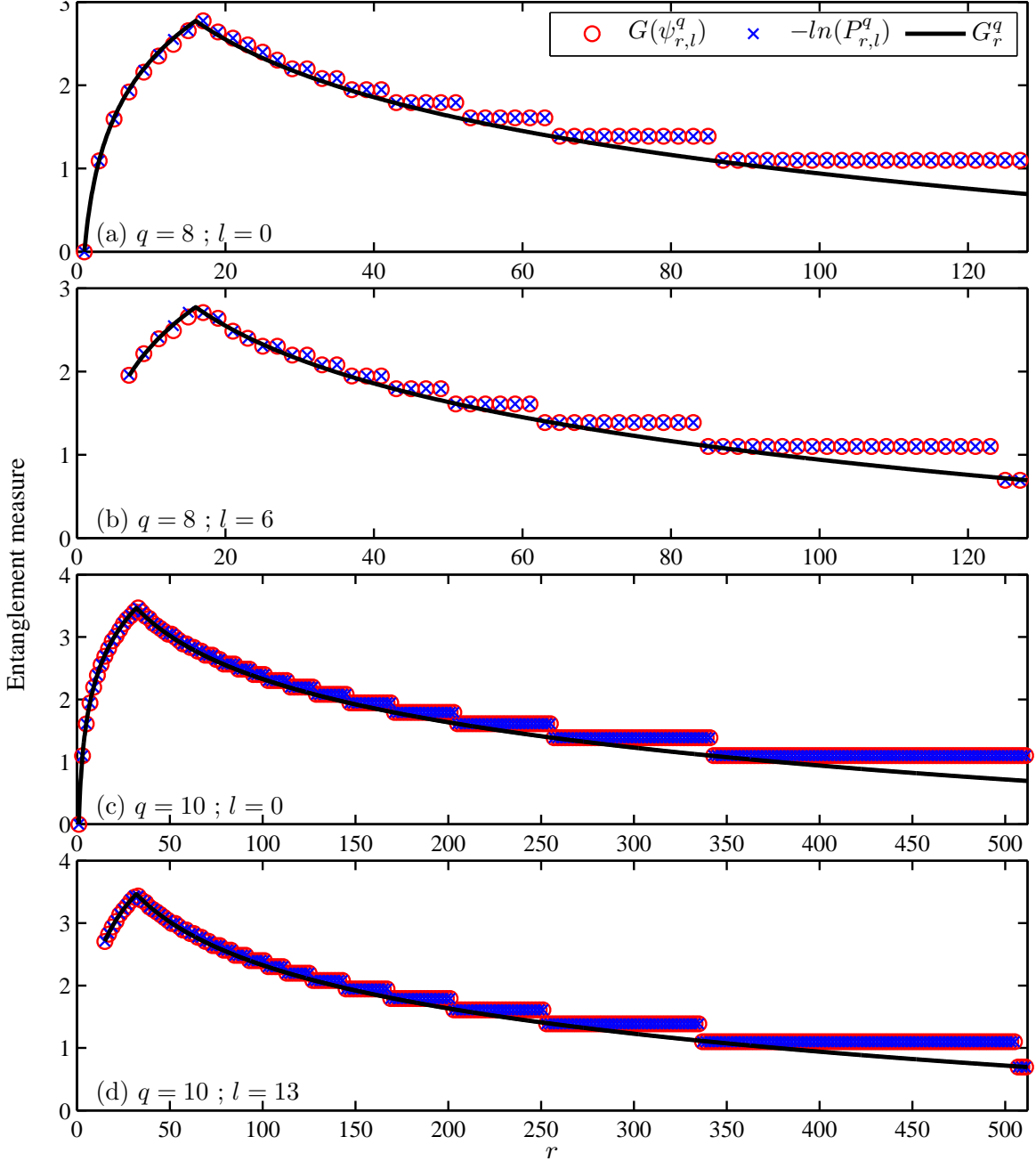}
\caption{(Color online)
Entanglement of periodic states as a function of their period, $r$,
for $q=8$ qubits and $l=0$ shift	
(a), $q=8$ and $l=6$ (b), $q=10$ and $l=0$ (c), and $q=10$ and
$l=13$ (d). The numerically calculated values are given in red
circles, the approximated (more accurate) formula is represented by blue crosses,
and the simple approximated (less accurate) 
formula is represented by a black line. 
Only odd periods were calculated. Similar results were
obtained with up to 12 qubits and all possible shifts.
}
\label{fig2}
\end{figure}

\noindent
Looking at $G^q_r$
as a function of $r$ 
(Fig.~\ref{fig2}) 
we see that
it consists of two branches:


(i) An {\it ascending branch}, which starts at $0$ when $r=1$, and rises to a
maximum at $r=\sqrt{Q}$. This branch covers the states with small
periods, which have a large number of basis states, and for which
the complete ES state is the presumed nearest product state.

(ii) A {\it descending branch}, which starts at the maximum and descents back
to $0$ as $r$ reaches $Q$. This branch covers the states with large 
periods, which have a small number of basis states, and for which a basis
state is the presumed nearest product state.


We shall now show evidence to support the validity of this approximation. 
We first note that in the case of an even period, the least
significant bit has the same value for all the basis states in the
superposition, which makes the state reducible. When factoring out
the last qubit, the rest of the qubits constitute an arithmetic
progression themselves, making them a periodic state with half the
period. The value of the last bit depends on the parity of the
shift:

\begin{equation}
|\psi^q_{2r,l}\rangle = |\psi^{q-1}_{r,\lfloor l/2 \rfloor}\rangle
\otimes |l \bmod 2\rangle.
\end{equation}

\noindent
Therefore, the problem of finding the entanglement of a periodic
state with an even period can be reduced to the same problem with a
state that has one less qubit and half the period. We shall concern
ourselves from now on only with states that have odd periods.
  
One important consequence of $r$ being odd is that $r$ is coprime to $2^m$
for every $m$. This means that the values of the $m$th significant bit in all
the basis states go through a $2^m$-long cycle with an equal number of
zeros and ones. If the number of cycles is whole then the total numbers
of zeros and ones in the $m$th bit are equal as well. However,
in most cases $2^m$ does not divide $A$, the cycle is truncated and the
equality is only approximate. Furthermore, the values of the $n$ least
significant bits make out an arithmetic progression with common
difference $r$ in respect to addition modulo $2^n$, which means they also 
go through all their possible values in a cyclic manner. Therefore, the 
values of the different bits are in general uncorrelated. This means 
that for any basis state $|k_0\rangle$, the number of states in $S$ that
have $n$ bits in common with it reduces approximately by half when $n$ 
is increased by 1. This is because for each qubit $m$, about half of all the
states have the same value for $m$ as $|k_0\rangle$, and the same is true
for any subset that is determined by the values of $n$ qubits. We conclude
that for a periodic state there is no basis state $|k_0\rangle$ that more
than any other basis state, has basis states in $S$ within a small 
Hamming distance from it. This again suggests that the approximated 
formula for $P_{\rm max}$ we have presented is valid for periodic states,
as opposed to the simple $W$ state.
  
An important observation concerning Eq.~(\ref{eq:pmaxperiodic}) is
that for the descending branch, where the nearest product state
$|\varphi\rangle$ is taken to be some basis state $k \in S$, it is at least
a local maximum of the overlap function $f^q_{r,l}$. To see this, note
first that there is no basis state $k' \in S$ that differs from $k$ 
on exactly one bit. If there was such a state, the difference $k-k'$
would be a power of 2, but this difference for a periodic state must be
a multiplicand of $r$, which is odd. For each $1\le m\le q$, the value
of $x_m$ is at the edge of its range (either $0$ or $1$), and trying to
vary its value will turn $|\varphi\rangle$ to a superposition of
$|k\rangle$ and $|k'\rangle$ which differ only on the $m$th bit.
Since $k' \notin S$ the overlap function will necessarily decrease, and
as this is so for all $m$ orthogonal directions, $|k\rangle$ is a local
maximum. The only question left then, for the descending branch, is
whether $|k\rangle$ is a global maximum as well.

Another justification to the approximated formula can be presented
in the form of an induction, using the following recursive
decomposition of periodic states. Considering the basis states 
that make up a periodic state, we can divide them into two subsets, 
according to the value of the most significant bit. Looking at the $q-1$
remaining bits, we see that each subset makes up a 
{\it component periodic state} with the same period:

\begin{equation}
|\psi^q_{r,l}\rangle = 
\sqrt{\frac{A_0}{A}} |0\rangle \otimes |\psi^{q-1}_{r,l}\rangle +
\sqrt{\frac{A_1}{A}} |1\rangle \otimes |\psi^{q-1}_{r,l'}\rangle.
\label{eq:recursive}
\end{equation}

\noindent
Here, $l'=-2^{q-1} (\bmod ~r)$ is the shift of the second component
periodic state, $A_0$ and $A_1$ are the number of basis states in
the component states $|\psi^{q-1}_{r,l}\rangle$ and
$|\psi^{q-1}_{r,l'}\rangle$ respectively. 
Clearly, the condition
$A=A_0+A_1$ is satisfied. 
When $A$ is even, $A_1=A_0=A/2$. 
When $A$ is odd $A_1=A_0-1$, 
namely $A_0=(A+1)/2$ and $A_1=(A-1)/2$. Let us now
assume that the approximated formula is correct for $q-1$ qubits.
If the states are in the ascending branch, then the complete ES
state is the approximated nearest product state of the component
states which make up the right-hand side of
Eq.~(\ref{eq:recursive}). Clearly this means that it is the nearest  
product state of the left-hand side as well, which is the periodic
state of $q$ qubits. On the other hand, if the states are in the
descending branch, each of the component states has  
a (different) basis state as the approximated nearest product
state. In this case, choosing one of these basis states gives the
nearest product state to the state of $q$ qubits.
  
Finally, we present numerical evidence to support the approximated formula,
described in Fig.~\ref{fig2}. This figure shows the
logarithmic Groverian entanglement of periodic states for some values of
$q$, $r$, and $l$, computed numerically via the numerical scheme
described in Sec. \ref{sec:groverian}. It also shows the
entanglement according to the approximated formula: both the 
more accurate
version given in Eq.~(\ref{eq:pmaxperiodic}) and the less accurate
version given in  Eq.~(\ref{eq:Gperiodic}). As shown in the
figure, the more accurate version of the formula agrees with the
numerical results to a good precision on both branches. The less
accurate version, however, does not follow the step-function-like
behavior in the descending branch. This is clearly due to the formula
for $A$, which includes a ceiling function that is smoothed out in
the approximation $\left\lceil (Q-l)/r \right\rceil \approx
Q/r$.
  
\section{Entanglement Induced by the QFT}
\label{sec:qft}

\begin{figure}
\includegraphics*[scale=0.49]{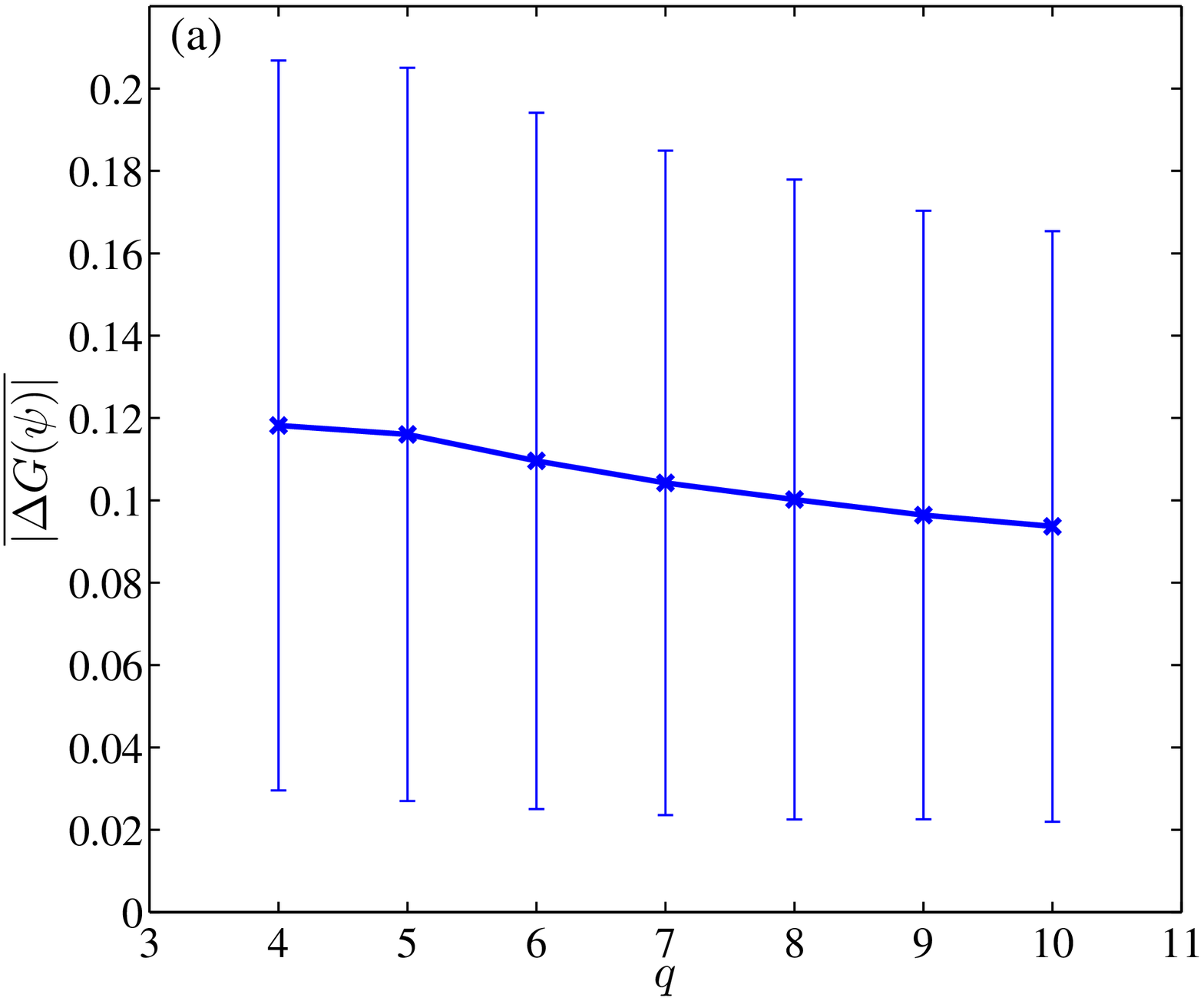}
\vspace*{-0.7cm}
\includegraphics*[scale=0.49]{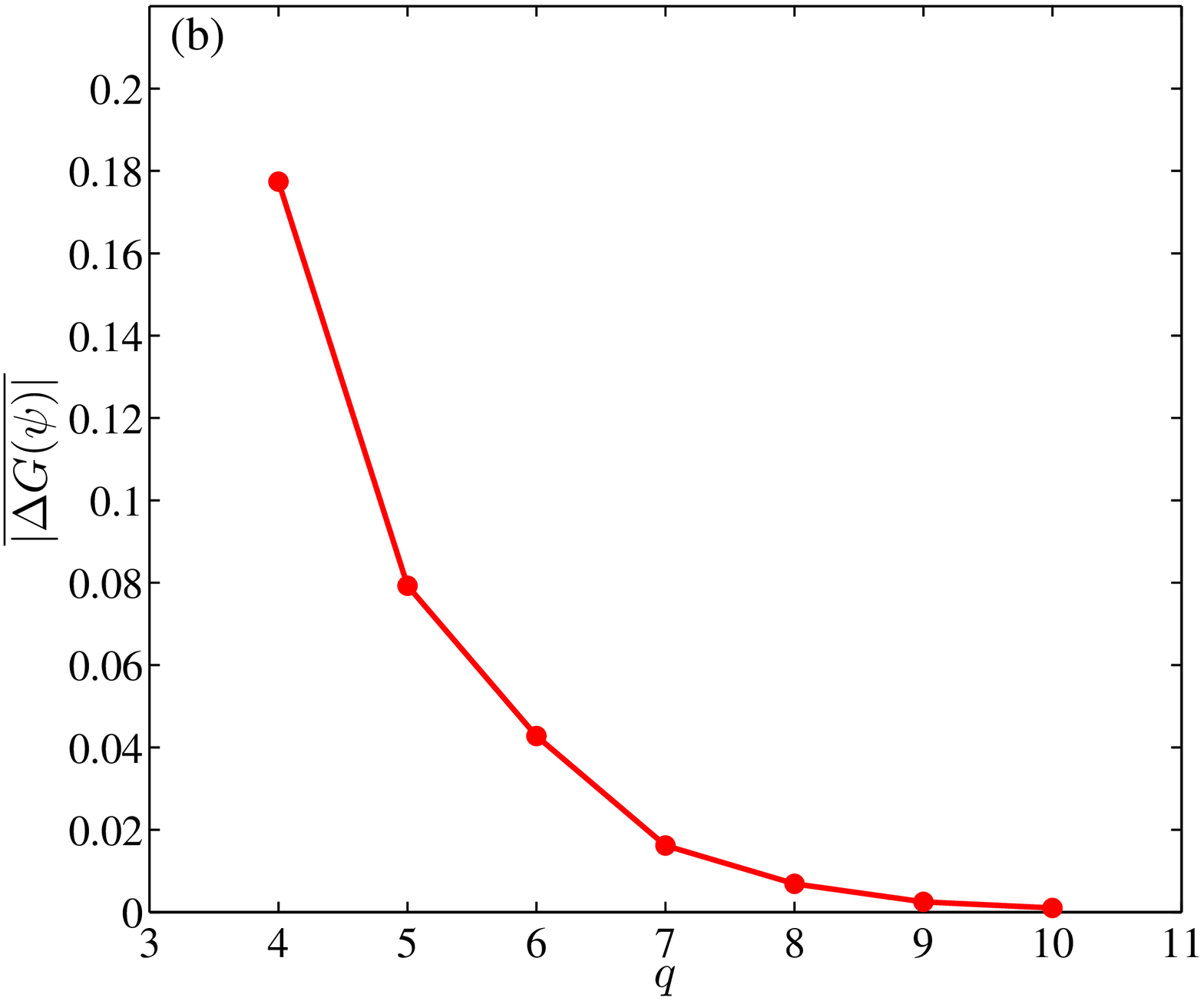}
\caption{(Color online)
The average absolute value of the change
$\Delta G(\psi)$ 
in the entanglement,
defined in Eq.~(\ref{eq:deltaG}), induced by the QFT
for a sample of random quantum
states (a) and for periodic states (b),
as a function of the number of qubits, $q$.
The random states are taken from
a uniform distribution on the $2q$-dimensional complex unit
sphere.
For the periodic states, the average is over
all periodic states with a given number of qubits.
For the random states
the error bars represent one standard deviation in
each direction, while for 
the periodic states
the distribution is extremely narrow, much narrower than the
width of the line.
}
\label{fig3}
\end{figure}

We have derived an approximated formula for the entanglement of periodic
states with the hope of better understanding why the QFT does not
increase their entanglement. First we note that this is indeed a special
property of periodic states, since as illustrated in Fig.~\ref{fig3},
the QFT operator in general changes the entanglement of quantum states.
Looking again at Eq.~(\ref{eq:qftperiodic}), we can see that the QFT
approximately takes a periodic state with period $r$ and shift $l$ to a
periodic state with period $Q/r$ and zero shift, up to relative
phases. This was shown more rigorously in
Eq.~(\ref{eq:exactqft}), where it was also shown that the relative phases
previously ignored take a very special form. To explain this, we
define a generalization of the regular ES state, by adding relative phases
that depend on a parameter $p$:  

\begin{equation}
|\psi^q_{S,p}\rangle = \frac{1}{\sqrt{ \left| S \right| }} 
\sum_{k \in S} e^{-\frac{pk}{Q}2\pi i} |k\rangle.
\end{equation}

\noindent
In this phased ES state the relative phase of each basis state is
proportional to the index of the state. An interesting property of this
state is that it can be obtained from the corresponding regular ES state
by local unitary operations, making it locally equivalent to the ES
state:

\begin{equation}
\left( {\begin{array}{*{20}c}
{1} & {0}  \\
{0} & {e^{-\frac{2^{q-1}p}{Q} 2\pi i}}  \\
\end{array}} \right) \otimes
\left( {\begin{array}{*{20}c}
{1} & {0}  \\
{0} & {e^{-\frac{2^{q-2}p}{Q} 2\pi i}}  \\
\end{array}} \right) \otimes
\dots \otimes
\left( {\begin{array}{*{20}c}
{1} & {0}  \\
{0} & {e^{-\frac{p}{Q} 2\pi i}}  \\
\end{array}} \right)
|\psi^q_S\rangle = |\psi^q_{S,p}\rangle
\end{equation}

\noindent
Naturally, this means that both states have the same entanglement. As
seen in Eq.~(\ref{eq:exactqft}), the relative phases added to the
periodic state $|\psi^q_{r,l}\rangle$ follow precisely the same
pattern, with $p=l+r(A-1)/2$. Therefore, they can be
ignored for all entanglement considerations.  
We are left with two changes the QFT induces in periodic states:
removing the shift and changing the period.
Interestingly, although the
importance of the QFT in Shor's
algorithm is in its canceling of the shift (which enables one to
extract the period), this change is irrelevant to the state's
entanglement, as the shift does not  
appear at all in the approximated formula.
Therefore we are left only with the 
change of period. Looking again at the approximated formula for the 
entanglement of periodic states given in 
Eq.~(\ref{eq:Gperiodic}), 
we see that the values of the two branches, $1/r$ and 
$r/Q$, are swapped by the operation $r \rightarrow Q/r$.
It is now clear why $G^q_r=G^q_{Q/r}$ and therefore why
$G(\psi^q_{r,l}) \approx G(\psi^q_{Q/r,0})$. 
The QFT operator takes each 
periodic state in the ascending branch to a corresponding periodic state in
the descending branch that has the same entanglement, and vice versa (recall
that the QFT operator is its own inverse).

Finally, we present numerical
evidence to support the claim that the QFT does not change the
entanglement of periodic states (for sufficiently large $q$).
To this end we define the change of the Groverian entanglement
of a state $\psi$ induced by the QFT

\begin{equation}
\Delta G(\psi) = G[{\rm QFT}(\psi)] - G(\psi).
\label{eq:deltaG}
\end{equation}

\noindent
We examine this difference for periodic states
as well as for random states.
The random states are taken from a 
uniform distribution on the $2q$-dimensional complex unit
sphere.
The Groverian measure of
random states of $q$ qubits exhibits
a distribution that was calculated before
for certain values of $q$ 
\cite{Most2007}.
Any unitary operator $U$ (such as the QFT),
when applied to a sample of these states,
will produce states whose Groverian measure exhibits the 
same distribution. Looking at values of
$\Delta G(\psi)$,
for some states it
is positive and for
other states it is negative.
The average of 
$\Delta G(\psi)$ 
over the random states is zero, but its distribution
exhibits a certain width.
In  
Fig.~\ref{fig3}
we show the average of the {\it absolute value} of 
$\Delta G(\psi)$,  
which is an estimate of the width of the distribution
of the values of
$\Delta G(\psi)$.
For periodic states
$\overline{|\Delta G(\psi)|}$  
quickly decreases to zero with increasing $q$ 
[Fig. \ref{fig3}(b)], 
while for random states
it changes only slightly
[Fig. \ref{fig3}(a)]. 
This demonstrates the special feature of periodic states,
namely, that their Groverian measure is not affected by the
QFT.

\section{Discussion}

The goal of the field of quantum algorithms is not only
to find quantum algorithms that present a speedup over classical
ones, but also to establish a deep understanding as to how this
speedup is made possible. Currently,
the best insight we can offer is that it is
made possible by the combination of quantum superposition and
quantum interference \cite{Feynman1982,Aharonov1999}. Quantum
superposition allows a sort of parallel computation, as all the
states in the superposition go through the same unitary evolution in
an independent manner. Since the results of each of the states
cannot be accessed directly, we cannot fully exploit quantum
parallelism. Nevertheless, through quantum interference the
different parallel paths can interact in a limited way, allowing
us to access some global properties of the resulting states. Finding
special situations where such a global property is the solution to
some computational problem (like the period in Shor's algorithm is
the solution to the factoring problem) is in fact the essence of quantum
algorithm design. The role of entanglement in this model of
quantum speedup is to allow quantum parallelism to reach its full
extent, since product states cover a very small range compared with
all possible superpositions 
\cite{Ekert1998}.
  
To exemplify this model for specific algorithms, one must try to
distinguish the role of quantum parallelism and quantum interference
in each algorithm. In the case of Shor's algorithm, a superposition
is built in the preprocessing stage, in a manner that does not make
use of interference (modular exponentiation is performed on each
computational basis state separately). This superposition in fact
already contains the desired information (the period), and the QFT
is merely needed to extract this information (by canceling the
shift). Therefore, it can be argued that the preprocessing stage is
where quantum parallelism is used, and the QFT introduces quantum
interference. This claim is supported by the fact that the QFT does
not increase the entanglement of the register.

\section{Summary}

We have shown that periodic states play an important role in Shor's
factoring algorithm and pointed out that this is also true for
other quantum algorithms that rely on the QFT. Focusing on
entanglement as a necessary resource for quantum speedup, we set as
our goal to explain the result presented in Ref.
\cite{Shimoni2005}, according to
which the QFT hardly affects the entanglement of periodic
states. For this purpose we analyzed the entanglement of periodic
states using the Groverian entanglement measure. We derived an
approximated formula for the Groverian entanglement of periodic
states and showed evidence to support it. Using this approximated
formula, we presented a model that explains the aforementioned
result. Finally, we argued that this result and the model that
explains it strengthen our understanding as to the source of quantum
computational speedup in Shor's algorithm and in general.

\clearpage
\newpage

\end{document}